\documentclass[12pt]{iopart}
\usepackage{iopams} 
\usepackage{epsfig}
\usepackage{exscale}
\usepackage{graphicx}
\usepackage{subfigure}	
\begin{document}

\newcommand{\La}{LaFeAsO$_{1-x}$F$_x\ $}
\newcommand{\LAO}{LaAlO$_3\ $}

\title{Epitaxial \La thin films grown by pulsed laser deposition}

\author{M. Kidszun, S. Haindl, E. Reich, J. H\"{a}nisch, L. Schultz and B. Holzapfel}

\address{IFW Dresden, Institute for Metallic Materials, P. O. Box 270116, D--01171 Dresden, Germany.}

\ead{M.Kidszun@ifw-dresden.de}

\begin{abstract}
Superconducting and epitaxially grown \La thin films were successfully prepared on (001)--oriented \LAO substrates using pulsed laser deposition. The prepared thin films show exclusively a single in--plane orientation with epitaxial relation (001)[100]$\|$(001)[100] and a FWHM value of 1$^{\circ}$. Furthermore, resistive measurement of the superconducting transition temperature revealed a $T_{\mathrm{c,90\%}}$ of 25$\:$K with a high residual resistive ratio of $6.8$. The applied preparation technique, standard thin film pulsed laser deposition at room temperature in combination with a subsequent post annealing process, is suitable for fabrication of high quality \La thin films. A high upper critical field of $76.2\:$T was evaluated for magnetic fields applied perpendicular to the \textit{c}--axis and the anisotropy was calculated to be 3.3 assuming single band superconductivity.
\end{abstract}

\maketitle

\section{Introduction}
\label{sec:Introduction}

The discovery of high temperature superconductivity in \textit{R}FeAsO$_{1-x}$F$_{x}$ (\textit{R} = rare earth) iron pnictides \cite{1,2} triggered the search for new superconductors and phenomena. Recently, further iron--based superconducting compounds were discovered \cite{3,4}. 
Concerning quaternary iron pnictides (`1111'--phase) most of the experimental investigation performed up to now used polycrystals or the available single crystals of NdFeAsO$_{1-x}$F$_{x}$ and SmFeAsO$_{1-x}$F$_{x}$ \cite{11,12}. 
Besides the challenge of single crystal growth also thin film fabrication of the fluorine doped oxypnictides has been addressed recently due to the importance of thin films for fundamental studies as well as for applications. Thin film fabrication of superconducting quaternary `1111'--compounds emerged to be very challenging \cite{13,14}. The first superconducting thin film has been grown by our group on \LAO substrate using pulsed laser deposition (PLD) \cite{15,16}. Difficulties and detailed investigation on the thin film growth of the La `1111'--phase is reported in Ref.~\cite{29}. To the best of our knowledge, there are no other reports on superconducting epitaxial thin films of the `1111'--phase. 

The fabrication of highly qualitative epitaxial grown thin films is mandatory for fundamental studies, e.g. on Josephson junctions, as well as for the development of multilayers or electronic devices. Furthermore, the deposition of textured thin films is so far the best alternative to the difficult single crystal growth. Due to their geometry and dimensionality, thin films open the beneficial way to a variety of experiments, such as transport current measurements, spectroscopy, multilayers, and possible new structures. Even though there are reports on single crystals of the \La phase \cite{17}, no detailed measurements of physical properties are published so far.

In this rapid communication we present the successful fabrication of epitaxial thin films of these new superconductors by pulsed laser deposition. Research using epitaxial grown \La thin films offers great opportunities to study the intrinsic properties of the new iron based superconductors. Structural as well as superconducting properties are discussed.

\section{Experimental details}
\label{sec:ThinFilmPreparation}

Thin films were deposited using a standard on--axis PLD setup with a KrF laser (Lambda Physik) with a wavelength  $\lambda = 248$\,nm, a pulse duration $\tau = 30$\,ns, and an energy density $\epsilon \approx 4$\,Jcm$^{-2}$ on the target surface. Vacuum conditions in the deposition chamber were $p = 10^{-6}$\,mbar. From our early results on thin film deposition in this system we expected a high fluorine loss during deposition. Hence, a target with a rather high fluorine content was used. The nominal target composition was LaFeAsO$_{0.75}$F$_{0.25}$ as calculated from the initial weight used for the target preparation.

The film presented in this publication was prepared using ex--situ phase formation. After thin film deposition using standard PLD at room temperature a 7$\:$h heat treatment at 960$\:^{\circ}$C was applied. Following film deposition and ex--situ phase formation structural properties were investigated via standard x-ray diffraction methods. The superconducting transition temperature $T_\mathrm c$ was measured in a commercial Physical Property Measurement System (PPMS Quantum Design) resistively in a four--point geometry up to $9\:$T.

\section{Results and discussion}
\label{sec:ResultsAndDiscussion}

\begin{figure}[htbp]
	\centering
		\includegraphics[width=0.50\textwidth]{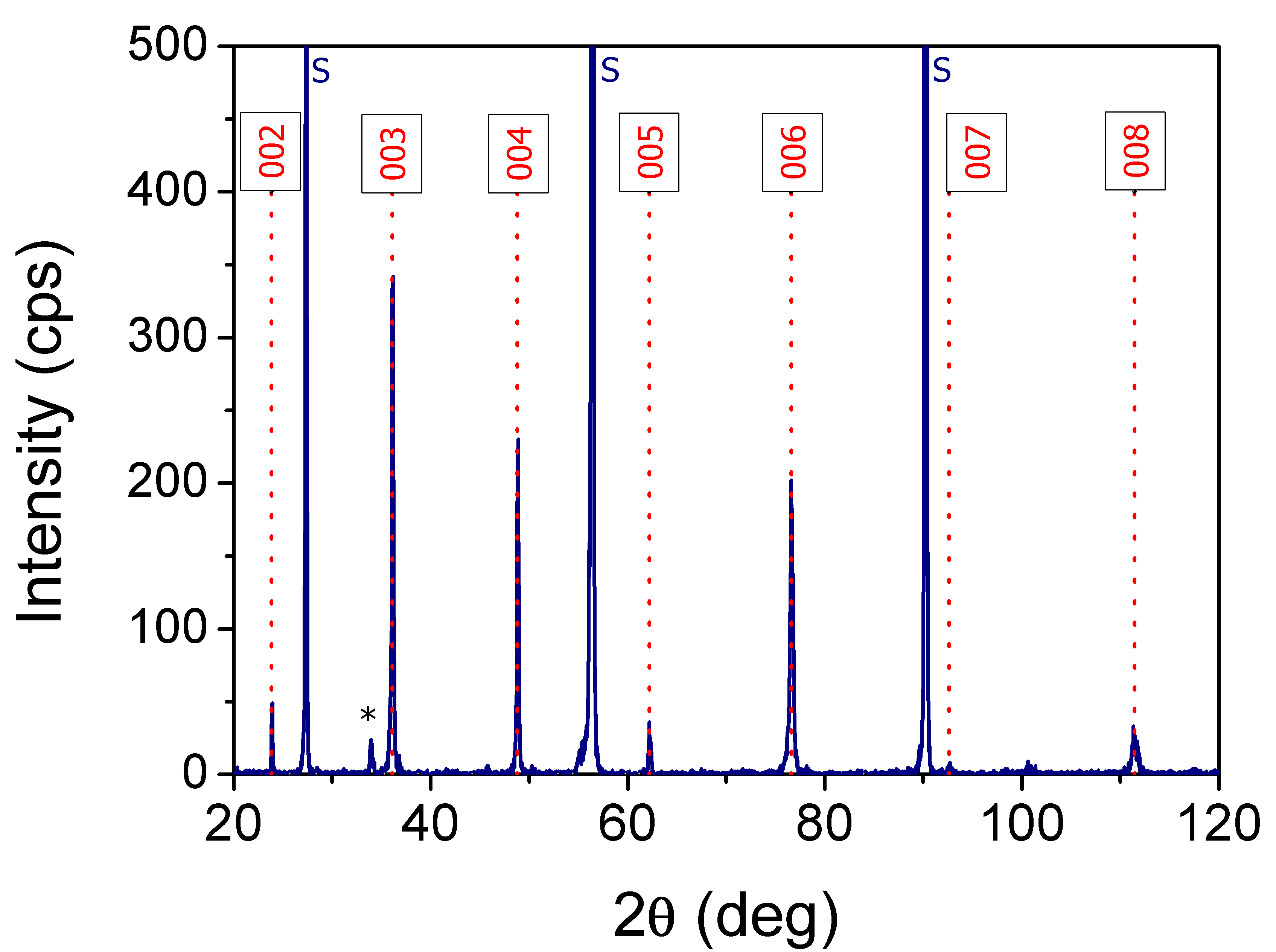}
	\caption{X--ray diffraction pattern of a fully epitaxial grown \La thin film. The \La phase shows exclusively $(00l)$ peaks. The substrate is marked with S and a small amount of oxide impurities is visible indexed by the black star.}
	\label{fig:Graph1}
\end{figure}

Detailed investigation of film growth and phase formation of the \La system was performed using different preparation methods \cite{15,29}. As result, a successful preparation technique was developed to fabricate epitaxial films grown on (001)--oriented \LAO single crystals as substrates. During the ex--situ annealing process the \La phase was formed. The x--ray diffraction pattern (Fig.~\ref{fig:Graph1}) revealed a c--axis textured \La thin film growth. Only $(00l)$--peaks of the \La phase are visible besides a small peak of LaOF, indicating nearly phase--pure \textit{c}--axis textured growth. As proven by TEM investigation on our polycrystalline thin films LaOF formation is a typical feature of this fabrication process. LaOF is formed at the interface between the substrate and the film and at the film surface \cite{29}. Oxygen impurities like LaOF also appeared during the single crystal growth as reported in reference \cite{17}. Optimising deposition condition and phase formation heat treatment, especially by reducing the oxygen partial pressures during film preparation and annealing, we succeeded in suppressing strongly the oxygen impurity phases resulting in an epitaxial film growth.

\begin{figure}[htbp]
	\centering
		\includegraphics[width=1.00\textwidth]{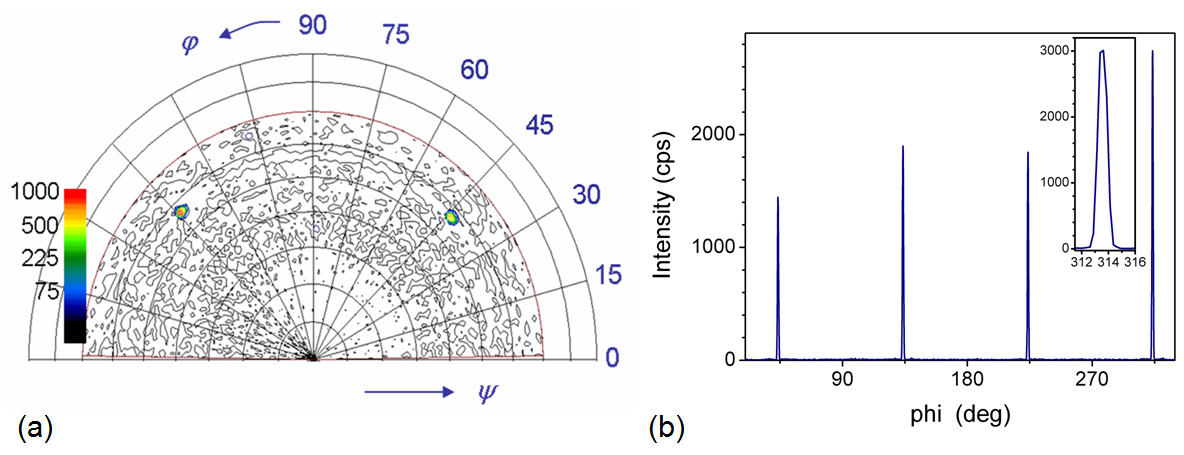}
	\caption{(a) X--ray (112) pole figure of a \La thin film grown epitaxially on \LAO substrate. (b) Phi--scan of the (112) peaks reaveals a sharp in--plane orientation distribution with a FWHM of 1$\:^{\circ}$.}
	\label{fig:polefigure}
\end{figure}

To check epitaxial growth of the \textit{c}--axis textured \La thin film (Fig.~\ref{fig:Graph1}), the (112) pole figure (Fig.~\ref{fig:polefigure}a) was measured.
In addition the (114),(102) and (314) pole figures were measured and considering the standard x--ray diffraction pattern (Bragg--Brentano) a P4/nmm crystal structure with lattice parameters of $c=0.8665\pm0.0005\:$nm and $a=b=0,407\pm0.003\:$nm was determined. This is a reasonable finding in fluorine doped \La ($x>0.1$) whereas the undoped compound has a lattice constant of $c=0.874\:$nm and $a=b=0,403\:$nm \cite{1}. As reported by Chen et al. \cite{31}, the fluorine doping reduces the \textit{c}--axis lattice constant.
The pole figures reveal a sharply textured in--plane aligned film. The in--plane orientation distribution of about 1$^{\circ}$ was determined by an extra phi--scan of the (112) poles depicted in Fig.~\ref{fig:polefigure}b. Furthermore, the phi scan indicates, that only one single epitaxial component is present with the epitaxial relationship (001)[100]$\|$(001)[100].

\begin{figure}[bp]
	\centering
		\includegraphics[width=1.00\textwidth]{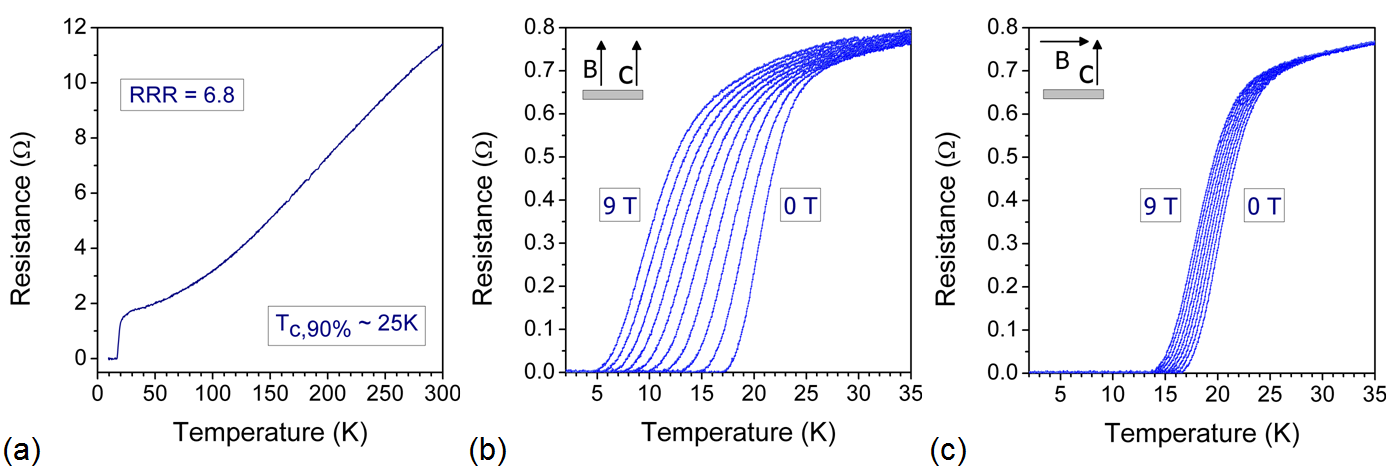}
	\caption{(a) Resistive measurement of the superconducting transition. $T_{\mathrm{c,90\%}}$ of this film is 25$\:$K whereas the film is fully superconducting at about 17$\:$K. Measurements in magnetic fields up to 9$\:$T in both major directions. (b) $B$ out--of--plane and (c) $B$ in--plane. }
		\label{fig:resistance}
\end{figure}

Resistive measurement results of the superconducting transition are shown in Fig.~\ref{fig:resistance}.
$T_{\mathrm{c,90\%}}$ of the epitaxial \La film is 25$\:$K whereas $T_{\mathrm{c}}(R\:$=$\:$0) is at 17$\:$K. 
The quality of the textured thin film emerges in the high residual resistive ratio ($RRR=6.8$). 
Figure~\ref{fig:resistance}b,c show the resistive behaviour of the thin film in an applied magnetic field up to 9$\:$T in both major directions: Fig.~\ref{fig:resistance}b shows out--of-plane (parallel to \textit{c}--axis) and Fig.~\ref{fig:resistance}c in--plane (perpendicular to \textit{c}--axis) measurements.

\begin{figure}[tbp]
	\centering
		\includegraphics[width=0.5\textwidth]{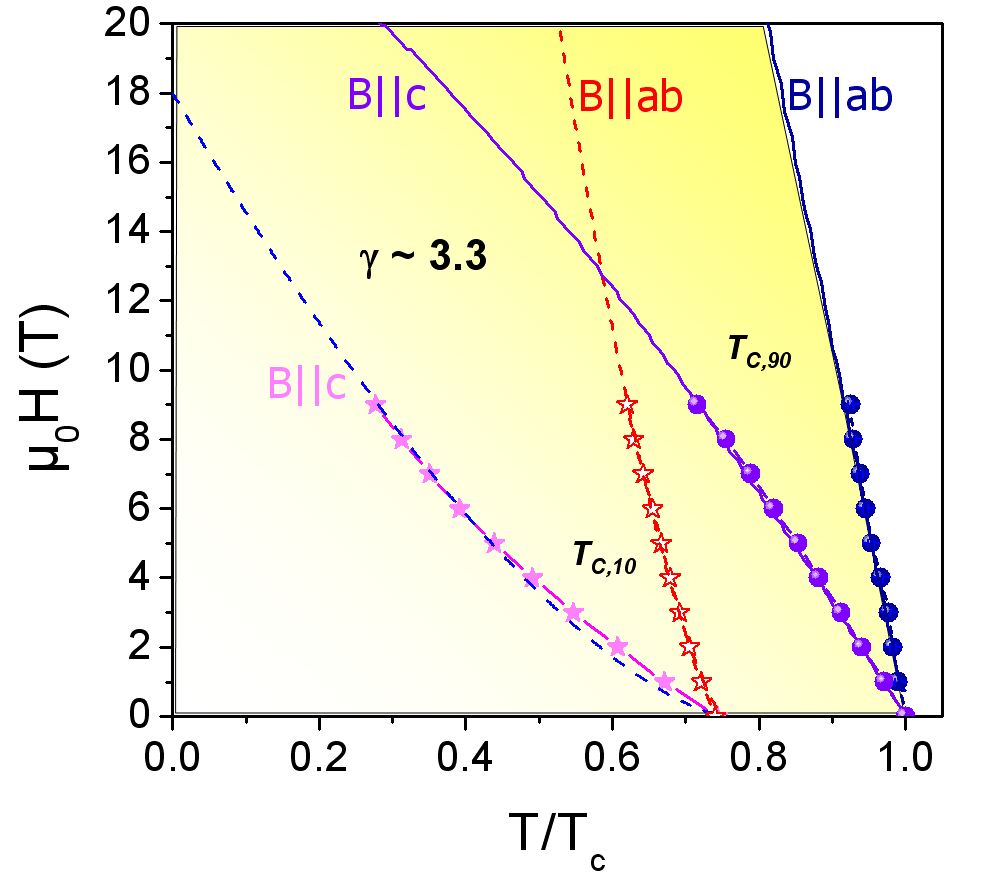}
	\caption{Magnetic phase diagram: $T_{\mathrm{c,90\%}}$ data with a fit based on WHH--single band model and $T_{\mathrm{c,10\%}}$ data fitted by a power law (exponent 1.5).}
	\label{fig:phasediagram}
\end{figure}

Based on $R(T)$ measurements, part of the magnetic phase diagram has been plotted (Fig.~\ref{fig:phasediagram}). The upper critical field, $\mu_0H_{\mathrm c2}(0)$, evaluated by the $T_{\mathrm{c,90\%}}$ criterion, i.e. 90\% of $R_{\mathrm{onset}}$, is approximated by the Werthamer--Helfand--Hohenberg (WHH) model for a single band ($\mu_0H_{\mathrm c2}(0)=0.693T_{\mathrm c}* | \frac{d\:\mu_0H_{\mathrm c2}}{d\:T} | _{T_{\mathrm c}}$). The irreversibility line, evaluated by the $T_{\mathrm{c,10\%}}$ criterion, is fitted using a power law with exponent 1.5. The upper critical field for magnetic fields applied parallel to the \textit{c}--axis was determined to be $\mu_0H_{\mathrm c2}^{\|}(0)=23.2\:$T with a slope 
$\frac{d\:\mu_0H_{\mathrm c2}^{\|}}{d\:T} | _{T_{\mathrm c}} = -1,34\:$T/K. In the case of an applied field perpendicular to the \textit{c}--axis, the upper critical field has a value of $\mu_0H_{\mathrm c2}^{\bot}(0)= 76.2\:$T  with a slope of $\frac{d\:\mu_0H_{\mathrm c2}^{\bot}}{d\:T} | _{T_{\mathrm c}} = -4.4\:$T/K.
Consequentially, the anisotropy factor was determined to be $\gamma=3.3$ based on single band calculations.

\section{Summary}
\label{sec:Summary}

In this study we report the first successful epitaxial growth of superconducting \La thin films with a $T_{\mathrm{c,90\%}}$ of 25$\:$K and a single in--plane orientation with FWHM of 1$^{\circ}$. The epitaxial relationship to the \LAO substrate is (001)[100]$\|$(001)[100]. We determined the anisotropy of the upper critical field to be 3.3 with a high $\mu_0H_{\mathrm c2}^{\bot}(0)= 76.2\:T$ perpendicular to the c--axis. The anisotropy was calculated with the assumption of single band superconductivity. Measurements on polycrystalline bulk \La ($x=0.11$), however, indicate a two--band behaviour \cite{32}. To investigate the full magnetic phase  diagram and clarify one or multi band behaviour high field measurements on this epitaxially grown thin films are required. 

\section*{Acknowledgement}
\label{sec:Acknowledgement}

The authors would like to thank F. Kurth and Dr. I. Kazumasa for their assistance; and especially J. Werner and M. Langer for target preparation and fruitful discussions. The authors acknowledge financial support by the German Research Foundation (DFG).

\section*{Reference}
\label{sec:Reference}

\end{document}